\documentclass[a4paper,11pt]{article}
\usepackage{pos}
\usepackage{slashed}

\newcommand{\be}{\begin{equation}}
	\newcommand{\ee}{\end{equation}}

\newcommand{\bea}{\begin{eqnarray}}
	\newcommand{\eea}{\end{eqnarray}}

\newcommand{\bei}{\begin{itemize}}
	\newcommand{\eei}{\end{itemize}}

\newcommand{\nn}{\nonumber}

\newcommand\eqn[1]     {eq.\,(\ref{#1})}
\newcommand\eqns[2]    {eqs.\,(\ref{#1}) and~(\ref{#2})}

\newcommand{\nm}{n_-}
\newcommand{\np}{n_+}

\newcommand{\T}{{\bf T}}

\newcommand{\as}{\alpha_s}
\newcommand{\eps}{\epsilon}
\newcommand{\ord}{{\cal O}}

\allowdisplaybreaks
\title{Factorization and 
resummation at next-to-leading-power}

\ShortTitle{Factorization and 
resummation at next-to-leading-power}

\author*[a]{Leonardo Vernazza}

\affiliation[a]{INFN, Sezione di Torino, Via P. Giuria 1, I-10125 Torino, Italy}

\emailAdd{leonardo.vernazza@to.infn.it}

\abstract{We discuss recent progress concerning the 
resummation of large logarithms at next-to-leading 
power (NLP) in scattering processes such as Drell-Yan 
and deep inelastic scattering near threshold, and 
thrust in the two-jet limit. We start by reviewing 
the approach based on soft-collinear effective field 
theory and show that the standard factorization into 
short distance coefficients, collinear and soft 
functions at NLP leads in general to the appearance 
of endpoint divergences, which prevent the naive 
application of resummation techniques based on the 
renormalization group. Taking thrust as a case 
study, we then show that these singularities 
are indeed an artifact of the effective theory, 
and discuss how they can be removed to recover 
a finite factorization theorem and achieve 
resummation at NLP, at LL accuracy. Last, 
we discuss recent work concerning the calculation 
of all collinear and soft functions necessary to 
reproduce Drell-Yan near threshold up to NNLO in 
perturbation theory. This calculation provides 
useful data to extend resummation at NLP beyond 
LL accuracy.}

\FullConference{16th International Symposium on Radiative Corrections: Applications of Quantum Field Theory to Phenomenology (
	RADCOR2023)\\
	28th May - 2nd June, 2023\\
	Crieff, Scotland, UK\\}

\begin{document}
\maketitle

\section{Soft-collinear radiation at NLP}

\begin{figure}[t]
	\centering
	\includegraphics[width=0.95\textwidth]{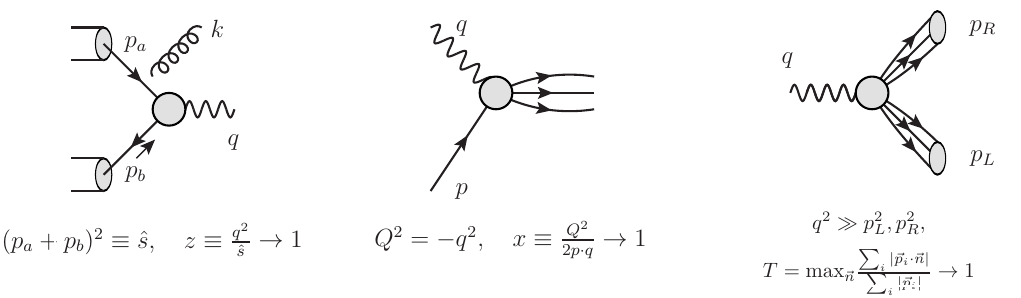}
	\caption{Kinematic definitions for the processes 
	considered in this talk.}
	\label{fig:figure1}
\end{figure}
In this talk we discuss Drell-Yan (DY) and deep 
inelastic scattering (DIS) near partonic threshold, 
and thrust in the two-jet limit (fig.~\ref{fig:figure1}). 
Defining $\hat s \equiv (p_a+p_b)^2$ the partonic 
centre of mass energy in DY and $Q^2 = q^2$ the 
invariant mass of the final state off-shell photon, 
the partonic threshold is defined by the condition 
$z \equiv Q^2/\hat s \to 1$; similarly, assigning 
momentum $p$ to the incoming parton in DIS, and 
defining $Q^2 = -q^2$ the invariant mass of the 
incoming photon, the threshold limit is given by 
the Bjorken variable $x \equiv Q^2/(2p\cdot q) \to 1$. 
Last, the two-jet limit in thrust is defined by 
the condition $T = {\rm max}_{\vec n} 
\frac{\sum_i |\vec p_i \cdot \vec n|}{\sum_i |\vec p_i|} \to 1$,
where $\vec{n}$ is the three-vector defining 
the thrust axis.
Labeling collectively the variables $z$, $x$, 
$T$ by $\xi$, in the limits above the partonic 
cross section is written as a power expansion 
in $(1-\xi)\to 0$, with each term developing 
towers of large logarithms in perturbation 
theory: 
\be
\hat{\sigma}(\xi) \sim \sum_{n=0}^{\infty}
\bigg(\frac{\as}{\pi} \bigg)^n \Bigg\{ 
c_{n} \delta(1-\xi) + \sum_{m=0}^{2n-1} 
\Bigg(c_{nm} \bigg[\frac{\ln^m(1-\xi)}{1-\xi}\bigg]_{+}
+ d_{nm} \ln^m(1-\xi) \Bigg) + \ord(1-\xi) \Bigg\}. 
\ee
In this equation the terms $c_{n}$ and 
$c_{nm}$ represent the leading power (LP) 
contribution, while the terms $d_{nm}$ give 
the next-to-leading power (NLP) correction.
The towers of large logarithms spoil the 
convergence of the perturbative series, 
and need to be resummed. For a long time 
it has been known how to resum the tower 
of logarithms in the LP term, see e.g. the 
seminal papers \cite{Parisi:1979xd,Curci:1979am,Sterman:1986aj,Catani:1989ne,Catani:1990rp}. 
Recently, a lot of effort has been devoted 
to the development of resummation for the 
towers of logarithms appearing at NLP. 
\begin{figure}[t]
	\centering
	\includegraphics[width=0.40\textwidth]{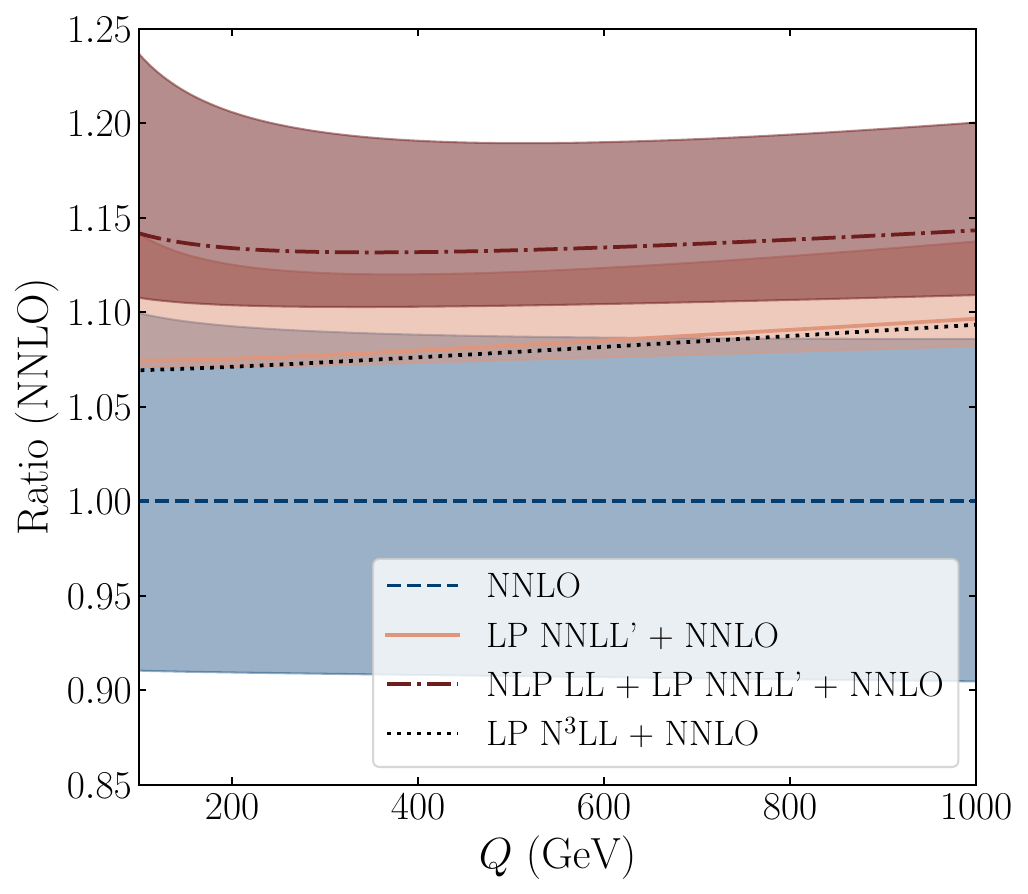}
	\hspace{0.5cm}
	\includegraphics[width=0.40\textwidth]{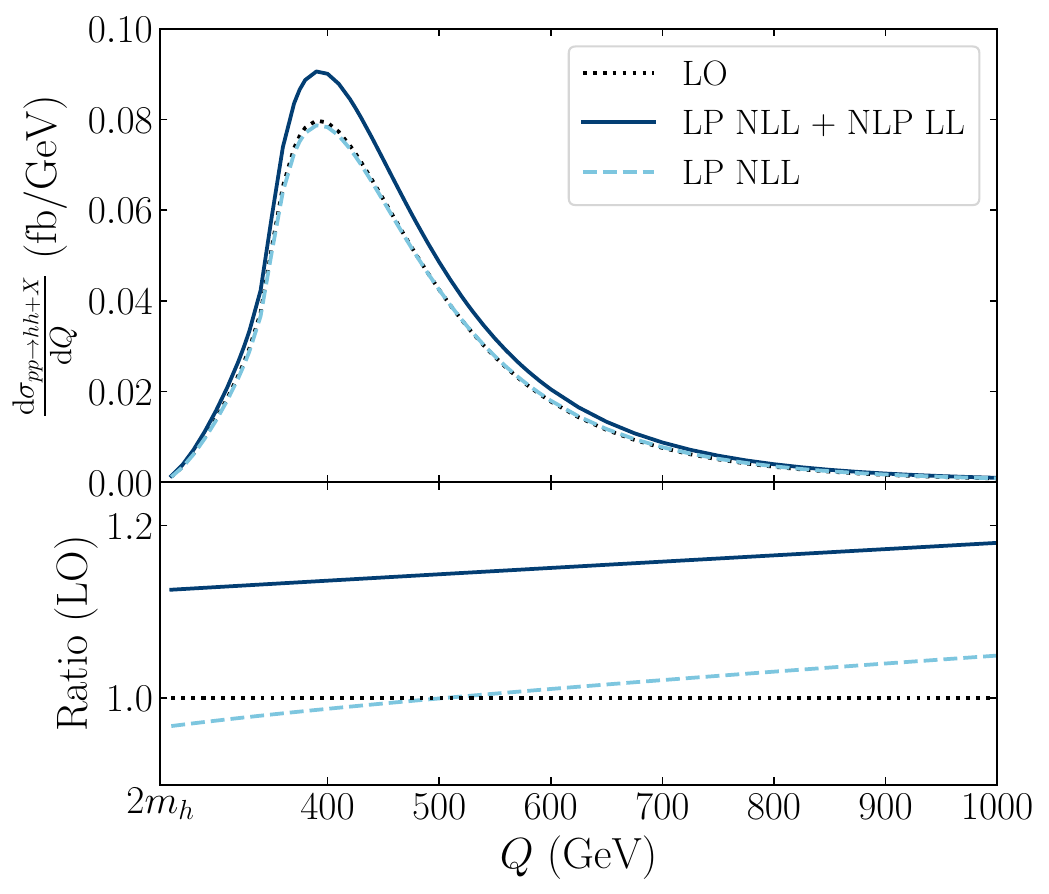}
	\caption{Numerical analysis showing the effect of including resummation of large 
		threshold logarithms at NLP on top of the LP component, in Higgs and di-Higgs 
		production, from \cite{vanBeekveld:2021hhv}.}
	\label{fig:figure2}
\end{figure}
It has been shown (see 
e.g.~\cite{Kramer:1996iq,Anastasiou:2014lda} 
and \cite{Beneke:2019mua,vanBeekveld:2021hhv}
for more recent analysis) that the 
resummation of logarithms at NLP 
may be important for precision physics. 
For instance, in case of processes such 
as Drell-Yan and Higgs production in gluon 
fusion, the resummation of threshold leading 
logarithms (LL) at NLP gives a contribution 
of the same order as the tower of 
next-to-next-to-leading logarithms 
(NNLL) at LP (fig.~\ref{fig:figure2}). 
Therefore, it would be recommendable for 
any analysis of particle scattering near 
threshold with resummation at NN(N)LL 
accuracy at LP, to include the summation 
of large logarithms at NLL accuracy
at NLP as well. 

The development of resummation of 
NLP logarithms has been investigated 
both within direct QCD \cite{Laenen:2008ux,Laenen:2008gt,Laenen:2010uz,Bonocore:2014wua,Bonocore:2015esa,Bonocore:2016awd,DelDuca:2017twk,Gervais:2017yxv,Bahjat-Abbas:2018hpv,Bahjat-Abbas:2019fqa,vanBeekveld:2019prq,vanBeekveld:2019cks,Laenen:2020nrt,Bonocore:2020xuj,Bonocore:2021qxh,Bonocore:2021cbv,vanBeekveld:2021hhv,vanBeekveld:2021mxn,Agarwal:2023fdk,Ajjath:2020ulr,Ajjath:2020sjk,Ajjath:2020lwb,Ajjath:2021pre,Ajjath:2021lvg,Ajjath:2021bbm,Ajjath:2022kyb,Engel:2021ccn,Engel:2023ifn,Czakon:2023tld}, 
and by means of effective field theory 
(EFT) methods based on soft-collinear 
effective field theory (SCET) \cite{Larkoski:2014bxa,Moult:2016fqy,Moult:2017rpl,Moult:2017jsg,Ebert:2018lzn,Moult:2018jjd,Moult:2019mog,Moult:2019uhz,Beneke:2017ztn,Beneke:2018rbh,Beneke:2018gvs,Beneke:2019kgv,Beneke:2019mua,Beneke:2019oqx,Beneke:2020ibj,Broggio:2021fnr,Beneke:2022obx,Broggio:2023pbu}. 
In this talk we will summarize recent 
developments within the second approach.

SCET \cite{Bauer:2000yr,Bauer:2001yt,Beneke:2002ph}
provides tools to describe soft and collinear 
radiation systematically, in principle at any 
subleading power. One introduces soft $q_s$, 
$A_s$ and collinear $\xi_c$, $A_c$ fields, 
that represent respectively soft and 
collinear modes of the original QCD 
fields. Hard modes are integrated out, 
and appear as short-distance (Wilson)
coefficients of effective operators. 
Let us start by recalling how 
factorization and resummation is 
achieved at LP. The effective operators 
describe the hard scattering kernel of 
a given process and are written in terms 
of gauge invariant fields, 
$\chi_c \equiv W_{c}^{\dag} \xi$ for 
quarks, ${\cal A}_c \equiv W_{c}^{\dag} 
[ i D_{c\perp}^{\mu} W_{c}^{\dag}]$
for gluons, where $W_{c}$ is a collinear
Wilson line, see e.g. eq.~(2.6) of \cite{Beneke:2019oqx}
for a definition. Each field is associated 
with one of the external particles. For 
instance, the processes in fig.~\ref{fig:figure1}
are all described at LP by the current 
\be\label{LPoperators}
\big[\bar \psi \gamma_{\mu} \psi\big](0) 
= \int dt \, d\bar t\, \widetilde{C}^{A0}(t,\bar t) 
\, J^{A0}_{\mu}(t,\bar t), 
\qquad \qquad 
J^{A0}_{\mu}(t,\bar t) = 
\bar \chi_{\bar c}(\bar t n_-) \, 
\gamma_{\perp \mu}\, \chi_{c}(t n_+), 
\ee
where $\widetilde{C}^{A0}(t,\bar t)$
is the Wilson coefficient in position 
space. Soft and collinear radiation 
arises in the effective theory by 
means of the SCET Lagrangian 
\be
{\cal L}_{\rm SCET} = 
\sum_i {\cal L}_{c_i} 
+ {\cal L}_{s}.
\ee
Collinear interactions occur within 
a given collinear sector by means 
of the collinear Lagrangian ${\cal L}_{c}$;
radiation among the different sectors can 
only be soft and involves the soft Lagrangian 
${\cal L}_{s}$. At LP soft-collinear interactions 
arise only due to a single term in 
${\cal L}^{(0)}_{c}$:
\be
{\cal L}^{(0)}_{c} = \bar{\xi}_c 
\bigg( i n_- D_c + g_s n_- A_s(x_-)
+i \slashed{D}_{\perp c} \frac{1}{i n_+ D_c} 
i \slashed{D}_{\perp c} \bigg) \frac{\slashed{n}_+}{2}
\xi_c + {\cal L}_{c,\rm YM}^{(0)},
\ee
and occur at position $x_-$ due to multipole 
expansion of the soft field in collinear 
interactions. Furthermore, only the component 
$n_- A_s$ appears, which leads to the well-known 
eikonal Feynman rule $\propto i g_s n_-^{\mu}$. 
As it turns out, this interaction can be removed 
from the Lagrangian by means of a soft-decoupling
transformation \cite{Bauer:2001yt}: 
\be\label{decoupling}
\xi(x) \to Y_{+}(x_-) \xi^{(0)}(x), 
\qquad \qquad 
A_c^{\mu}(x) \to Y_{+}(x_-) A_c^{(0)\mu}(x) Y_{+}^{\dag}(x_-), 
\ee
where $Y_{+}$ is a soft Wilson line, 
see e.g. eq.~(2.4) of \cite{Beneke:2019oqx}, 
such that one has 
\be
\bar{\xi}_c 
\Big( i n_- D_c + g_s n_- A_s(x_-) \Big) 
\frac{\slashed{n}_+}{2} \xi_c 
\to \bar{\xi}^{(0)}_c 
\Big( i n_- D^{(0)}_c \Big) 
\frac{\slashed{n}_+}{2} \xi^{(0)}_c.
\ee
This construction guarantees the automatic 
factorization of a given matrix element
(or cross section) into a product of 
short distance coefficients times 
collinear and soft functions, defined 
as matrix elements of gauge-invariant 
operators made exclusively of collinear 
and soft fields respectively. In case 
of the processes in fig.~\ref{fig:figure1}
one obtains the factorized expression
\be\label{LPDY}
\frac{d\sigma}{dQ^2} = |C^{A0}(Q^2)|^2 
\times f_{a/A} \otimes f_{b/B} 
\otimes S_{\rm DY}\big[Q(1-z)\big], 
\ee
for the Drell-Yan invariant mass 
distribution \cite{Becher:2007ty}, 
\be\label{LPDIS}
F_2 = |C^{A0}(Q^2)|^2 \times Q^2 
\times f_{a/A} \otimes J_{\overline{hc}}^{(q)}\big[Q(1-z)\big], 
\ee
for Deep inelastic scattering 
\cite{Becher:2006mr}, and
\be\label{LPTH}
\frac{d\sigma}{d\tau} = |C^{A0}(Q^2)|^2 
\times J_{c}^{(q)}(p_L^2) \otimes J_{\bar c}^{(q)}(p_R^2) 
\otimes S_{\rm LP}(k), 
\ee
for thrust \cite{Becher:2008cf}, with 
$\tau = 1-T$ and $k$ a soft momentum, 
reproducing older results in QCD 
\cite{Sterman:1986aj,Catani:1989ne,Catani:1992ua}. 
In these equations the matrix elements 
of (anti-)collinear fields are interpreted 
either in terms of the parton distribution 
functions (DY and DIS) or jet functions, 
(DIS and Thrust), while the soft functions
are given as vacuum expectation values
of the soft Wilson lines introduced in 
\eqn{decoupling}. One of the most important 
features of the EFT approach is that the original 
infrared singularities of the scattering amplitude 
are turned into ultraviolet divergences of the 
EFT's operators \cite{Becher:2009qa}. The 
renormalization of such operators provides 
renormalization group equations (RGEs), whose 
solution allows one to sum large logarithms 
associated to the hard, soft and collinear 
functions, thanks to the fact that within 
the EFT factorization, these functions are 
single scale objects. 

Let's now consider the extension of this 
framework beyond LP. In this respect, one of 
the advantages of the EFT approach is that 
every object (fields, derivatives, momenta)
has a unique scaling with the small parameter 
in the problem, conventionally indicated by 
$\lambda \ll 1$. For instance, in case of 
Drell-Yan near threshold $\lambda \sim \sqrt{1-z}$. 
Decomposing momenta along the light-like 
directions $n_{i\pm}$, such that 
$p^{\mu} = (n_+p) n_-^{\mu}/2 + p_{\perp}^{\mu}
+ (n_-p) n_+^{\mu}/2 = (n_+p, p_{\perp}, n_-p)$, 
collinear and soft momenta have respectively 
scaling $p_c \sim Q(1,\lambda, \lambda^2)$ 
and $p_s \sim Q(\lambda^2,\lambda^2, \lambda^2)$.
In this context, one needs to take into account 
two sources of power suppression 
\cite{Larkoski:2014bxa,Beneke:2017ztn,Beneke:2018rbh}. 
On the one hand, one has operators that are 
power suppressed compared to the LP ones in 
\eqn{LPoperators}. In general, power 
suppression is achieved either by inserting 
transverse derivatives, $\partial_{\perp} 
\sim \lambda$, or by adding more collinear 
fields along the same collinear direction. 
It is also possible to insert 
gauge-invariant combinations of soft fields,
but these contribute in general beyond NLP.
Starting from the LP current in 
\eqn{LPoperators} we have e.g. 
\bea\label{NLPop} \nn
\bar \chi_{\bar c}(\bar t n_-) \, 
\gamma_{\perp}^{\mu}\, \chi_{c}(t n_+) \,\,&\to&\,\, 
\bar \chi_{\bar c}(\bar t n_-)   
\big[ n_{\pm}^{\mu} \, i \slashed{\partial}_{\perp} \big] \chi_{c}(t n_+), 
\hspace{1.75cm} \mbox{(A1-type),} \\ 
\,\,&\searrow& \,\,\bar \chi_{\bar c}(\bar t n_-)   
\big[ n_{\pm}^{\mu} \, \slashed{\cal A}_{c\perp} (t_2 n_+)\big] \chi_{c}(t_1 n_+),
\hspace{0.5cm} \mbox{(B1-type),}
\eea
where for the labeling of power suppressed
operators we refer to \cite{Beneke:2017ztn,Beneke:2018rbh}.
The operators on the r.h.s of \eqn{NLPop} are suppressed
by one power of $\lambda$ compared to the operator on the 
l.h.s. Given that $\lambda \sim \sqrt{1-z}$, in general 
one needs to take into account operators suppressed up 
to two powers of $\lambda$, in order to reproduce a given 
cross section up to NLP. The second source of power 
suppression originates by considering time-ordered 
products of LP operators with power-suppressed 
insertions of terms from the SCET Lagrangian. 
Given the collinear SCET Lagrangian ${\cal L}_{c} 
= {\cal L}^{(0)}_{c} + {\cal L}^{(1)}_{c}
+ {\cal L}^{(2)}_{c} + \ord(\lambda^3)$, 
one has e.g. two terms contributing to 
${\cal L}^{(1)}_{c}$ \cite{Beneke:2002ni}, 
given by 
\be\label{Lc1}
{\cal L}^{(1)\rm gluon}_{c} = 
\bar \xi \Big[ x_{\perp}^{\mu} n_-^{\nu} 
W_c g_s F_{\mu\nu}^s(x_-) W_{c}^{\dag} \Big] \frac{\slashed{n_+}}{2} \xi, 
\qquad
{\cal L}^{(1)\rm quark}_{c} =
\bar q(x_-) W_c^{\dag} i \slashed{D}_{\perp c} \xi,
\ee
where the first term involves the emission 
of a soft gluon and in the second a collinear 
quark is converted into a soft quark, upon 
emission of a collinear gluon. Now, the 
difference compared to our previous discussion 
of factorization at LP is that soft-collinear 
interactions at subleading power, such as 
the one in \eqn{Lc1}, are
not removed by the decoupling transformation 
\eqn{decoupling}. Formally it is still possible 
to proceed with the factorization of a given 
matrix element into its soft and collinear 
components. However, a few differences arise 
compared to the factorization theorems at LP. 
\begin{figure}[t]
	\centering
	\includegraphics[width=0.90\textwidth]{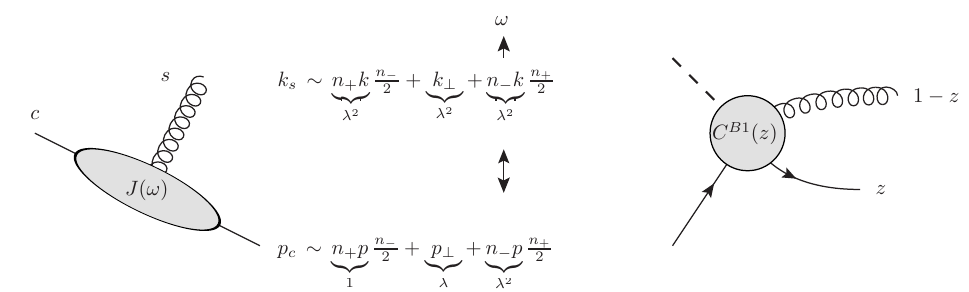}
	\caption{Convolutions appearing at NLP between jet and 
	soft functions (left, see e.g. \cite{Beneke:2018gvs,Beneke:2019oqx}) 
	or between short-distance coefficients and jet functions 
	(right, see e.g. \cite{Beneke:2020ibj}).}
	\label{fig:figure3}
\end{figure}
When the power suppression is given by a 
soft-collinear Lagrangian insertion such 
as in \eqn{Lc1}, we obtain 
a convolution between a collinear and a 
soft function, where the convolution variable 
is related to the small component of the 
collinear momentum, which is of the same order 
of the corresponding component of the soft 
momentum, (fig.~\ref{fig:figure3}, left 
diagram), and therefore cannot be 
integrated out. This factorization 
structure appear for instance in 
Drell-Yan \cite{Beneke:2018gvs,Beneke:2019oqx}.
When the power suppression is given by the 
insertion of an operator involving two or 
more particle in the same collinear sector, 
such as the B1 type operator in \eqn{NLPop}, 
convolution between the corresponding 
short-distance coefficient and jet 
function arises, where the convolution 
variable is related to the fraction 
of collinear momentum shared between 
the two particle in the same collinear 
sector (fig.~\ref{fig:figure3}, 
right diagram). This factorization 
structure arise for instance in
off-diagonal DIS \cite{Beneke:2020ibj}. 

\section{Endpoint divergences at NLP}

The presence of convolutions does not 
constitute a problem per se. However, 
as it turns out, these integrations are 
often divergent in $d = 4$. For instance, 
in case of DY one finds \cite{Beneke:2019oqx} 
\be\label{endpointDY}
\int d \omega \, J(\omega) \, S(\omega) 
\sim  \int_0^{\Omega} d \omega\,
\underbrace{(n_+ p \omega)^{-\eps}}_{\rm 
collinear \,\,piece} \underbrace{\frac{1}{\omega^{1+\eps}} 
\frac{1}{(\Omega - \omega)^{\eps}}}_{\rm soft \,\,piece}\bigg|_{\Omega=Q(1-z)}, 
\ee
which for $\eps \to 0$ is divergent for 
$\omega \to 0$. In case of off-diagonal 
scalar DIS \cite{Beneke:2020ibj} one 
finds
\be \label{endpointDIS}
\int d z \, C^{B1}(z) \, J_{qg}(z) 
\sim  \int_0^1 d z\,
\underbrace{\bigg(\frac{\mu^2}{s_{qg}z \bar z}\bigg)^{\eps}}_{\rm 
collinear \,\,piece} \underbrace{\frac{\alpha_s C_F}{2\pi}
\frac{(1-z)^2}{z}}_{\rm hard \,\,piece}\bigg|_{s_{qg}=Q^2\frac{1-x}{x}}, 
\ee
which for $\eps \to 0$ diverges for $z \to 0$. 
In order to investigate the structure of these endpoint 
divergences, let's consider the case of DIS more in detail:
In \eqn{endpointDIS} we have inserted the Wilson coefficient
at tree level. At one loop one has \cite{Beneke:2020ibj} 
\bea\label{CB11loop}\nn
C^{B1}(z)\Big|_{\rm 1\,\,loop} &\sim& 
C^{B1}(z)\Big|_{\rm tree} \frac{\as}{\pi}\frac{1}{\eps^2}
\bigg\{
{\bf T}_1\cdot {\bf T}_0 \bigg(\frac{\mu^2}{zQ^2}\bigg)^{\eps}
+{\bf T}_2\cdot {\bf T}_0 \bigg(\frac{\mu^2}{\bar zQ^2}\bigg)^{\eps}\\
&&\hspace{2.5cm}+\,
{\bf T}_1\cdot {\bf T}_2 \bigg[\bigg(\frac{\mu^2}{Q^2}\bigg)^{\eps} 
-\bigg(\frac{\mu^2}{zQ^2}\bigg)^{\eps}\bigg] \bigg\}
+ \ord(\eps^{-1})
\eea
The term $\propto {\bf T}_1\cdot {\bf T}_2$ contains a 
single pole, which however gives rise to a leading pole 
after integration. The correct result is obtained only 
within dimensional regularization:
\be
\frac{1}{\eps^2} \int_0^1 dz \,  
\frac{1}{z^{1+\eps}}(1-z^{-\eps}) = -\frac{1}{2\eps^3}, 
\ee
while expanding for $\eps \to 0$ leads to
nonsense results: 
\be
\frac{1}{\eps^2} \int_0^1 dz \,  
\frac{1}{z^{1+\eps}}\bigg(\eps \ln z 
- \frac{\eps^2}{2!}\ln^2 z 
+ \frac{\eps^3}{3!}\ln^3 z + \ldots\bigg)
= -\frac{1}{\eps^3} + \frac{1}{\eps^3}
- \frac{1}{\eps^3} + \ldots. 
\ee
This poses a problem for standard 
resummation techniques. As discussed 
above, within an EFT approach one 
renormalizes the collinear and soft 
matrix elements, obtaining a set of 
RGEs whose solution resums the large 
logarithms. From 
eqs.~(\ref{endpointDY}),~(\ref{endpointDIS}) 
and~(\ref{CB11loop}), however, it is 
clear that not all logarithms are generated 
within the hard, soft and jet functions
themselves: an additional pole (and thus 
a corresponding logarithm) is generated
through the endpoint divergent convolution. 
Closer inspection of \eqn{CB11loop} reveals 
that the endpoint divergence actually 
points a break of the EFT itself 
(fig.~\ref{fig:figure4}):
\begin{figure}[t]
	\centering
	\includegraphics[width=0.90\textwidth]{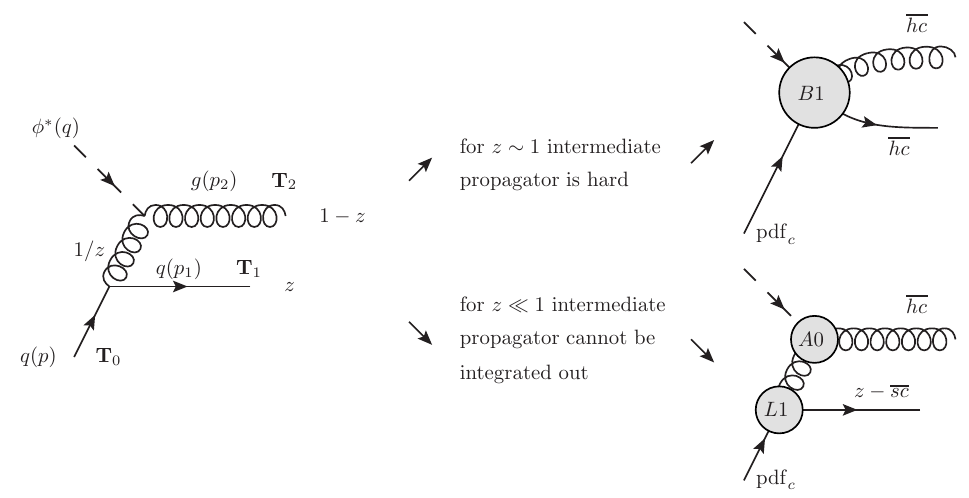}
	\caption{Structure of endpoint 
	divergence in off-diagonal scalar DIS, from \cite{Beneke:2020ibj}.}
	\label{fig:figure4}
\end{figure}
The factor $1/z$ in the Wilson 
coefficient $C^{B1}$ is due to 
the intermediate gluon propagator 
(l.h.s. of fig.~\ref{fig:figure4}).
For generic $z\sim 1$, this propagator 
is hard, thus an EFT description in terms
of a short-distance Wilson coefficient 
is appropriate (upper r.h.s. of 
fig.~\ref{fig:figure4}). However, $z$ 
is integrated in the range (0,1), and 
when $z\ll 1$, the intermediate 
propagator is not hard, and cannot 
be integrated out: the correct EFT
description is now given by the 
lower diagram on the r.h.s. of 
fig.~\ref{fig:figure4}. For $z\ll 1$
the short-distance coefficient $C^{B1}$
becomes a two-scale object, and together 
with the corresponding power-suppressed 
operator it \emph{refactorizes} into a 
jet function times the LP Wilson 
coefficient and operator $C^{A0}$
\cite{Beneke:2020ibj}:
\be
C^{B1}(Q,z) J^{B1} \stackrel{z\to 0}{\longrightarrow}
C^{A0}(Q^2) \int d^4 x \, T \Big[ J^{A0},
{\cal L}^{(1)\rm quark}_{z-\overline{sc}} \Big]
= C^{A0}(Q^2)D^{B1}(zQ^2) J^{B1}_{z-\overline{sc}},
\ee
where on the r.h.s.~$C^{A0}(Q^2)$ and 
$D^{B1}(zQ^2)$ can now be interpreted 
correctly as single-scale function. 
Such refactorization has been observed 
also in other applications of SCET 
to the analysis of processes at NLP, 
such as $H \to \gamma\gamma$, $H \to gg$ 
\cite{Liu:2019oav,Liu:2020tzd,Liu:2020wbn,Liu:2022ajh}, 
or in $B$-physics \cite{Boer:2018mgl,Feldmann:2022ixt,Cornella:2022ubo,Hurth:2023paz}
and muon-electron backward scattering \cite{Bell:2022ott}.

The analysis of DIS shows that a correct 
EFT treatment needs to take into account 
both the configurations appearing on the 
r.h.s.~of fig.~\ref{fig:figure4}. Once 
both contributions are taken into account, 
one expects endpoint divergences to cancel
between the two terms. In case of DIS, such
construction involves the factorization of 
the perturbative part of the initial state 
PDF, which makes the construction more 
involved. In the next section we will 
focus instead on off-diagonal thrust, 
where the cancellation of endpoint 
divergences can be shown explicitly,
without involving initial-state 
singularities.

\section{NLP LLs in Thrust in the two-jet limit}

\begin{figure}[t]
	\centering
	\includegraphics[width=0.24\textwidth]{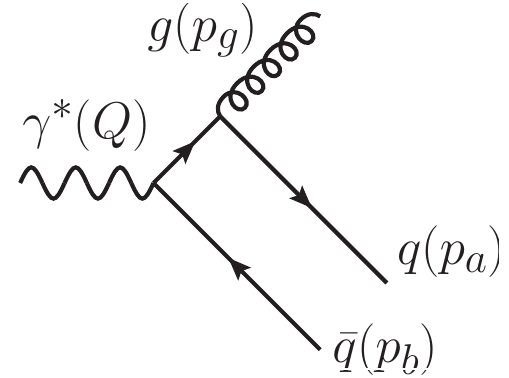}
	\hspace{3.0cm}
	\includegraphics[width=0.24\textwidth]{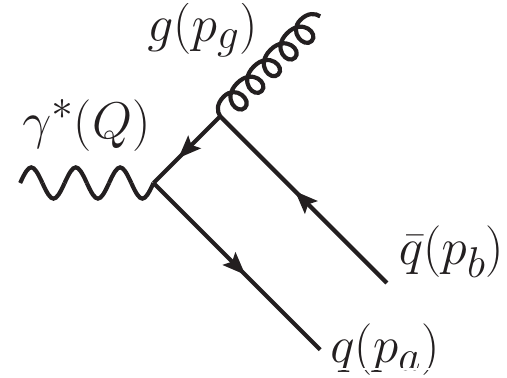}
	\caption{Off-diagonal ``gluon'' 
		thrust in the two-jet limit. Figure from \cite{Beneke:2022obx}.}
	\label{fig:figure5}
\end{figure}

\begin{figure}[t]
	\centering
	\includegraphics[width=0.26\textwidth]{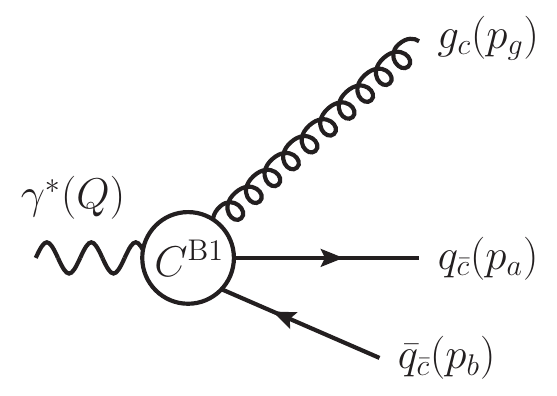}
	\hspace{1.0cm}
	\includegraphics[width=0.26\textwidth]{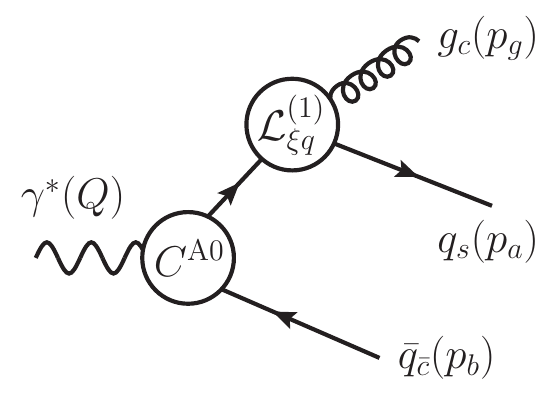}
	\hspace{0.2cm}
	\includegraphics[width=0.26\textwidth]{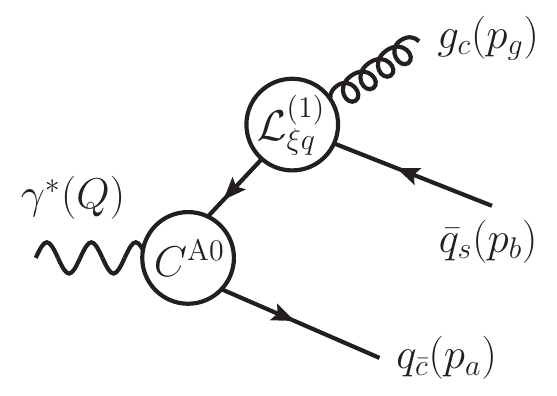}
	\caption{Matrix elements in SCET corresponding 
		respectively to \eqn{eq:btypeME} (first diagram 
		on the left) and \eqn{eq:atypetproduct} (second
		and third diagrams). Diagrams from \cite{Beneke:2022obx}.}
	\label{fig:figure6}
\end{figure}

\begin{figure}[t]
\centering
\includegraphics[width=0.95\textwidth]{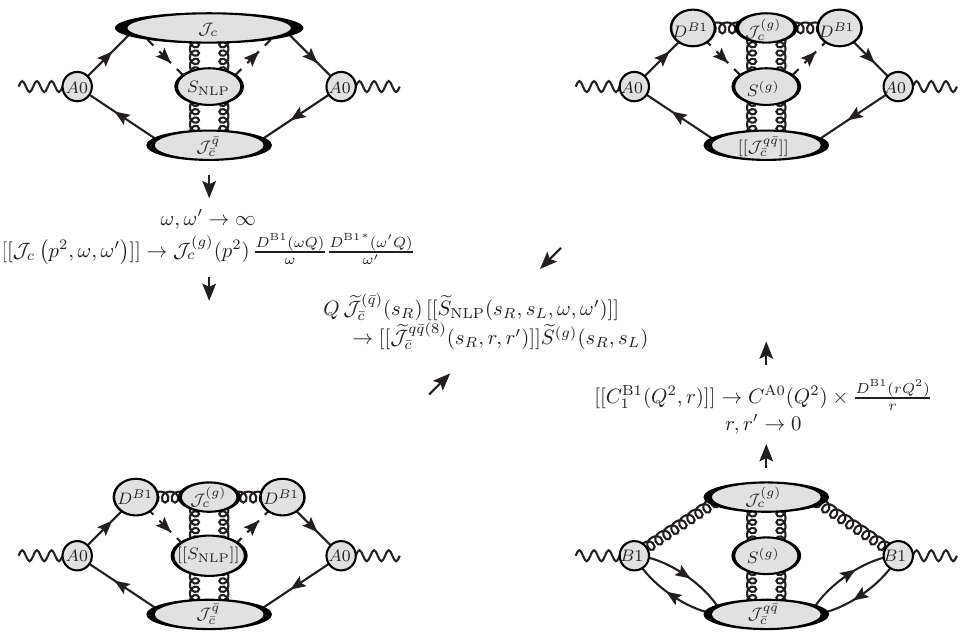}
\caption{Consistency conditions necessary to achieve 
	refactorization and the cancellation of endpoint 
	divergences, as obtained in \cite{Beneke:2022obx}.}
\label{fig:figure7}
\end{figure}

\begin{figure}
\begin{center}
\includegraphics[width=0.30\textwidth]{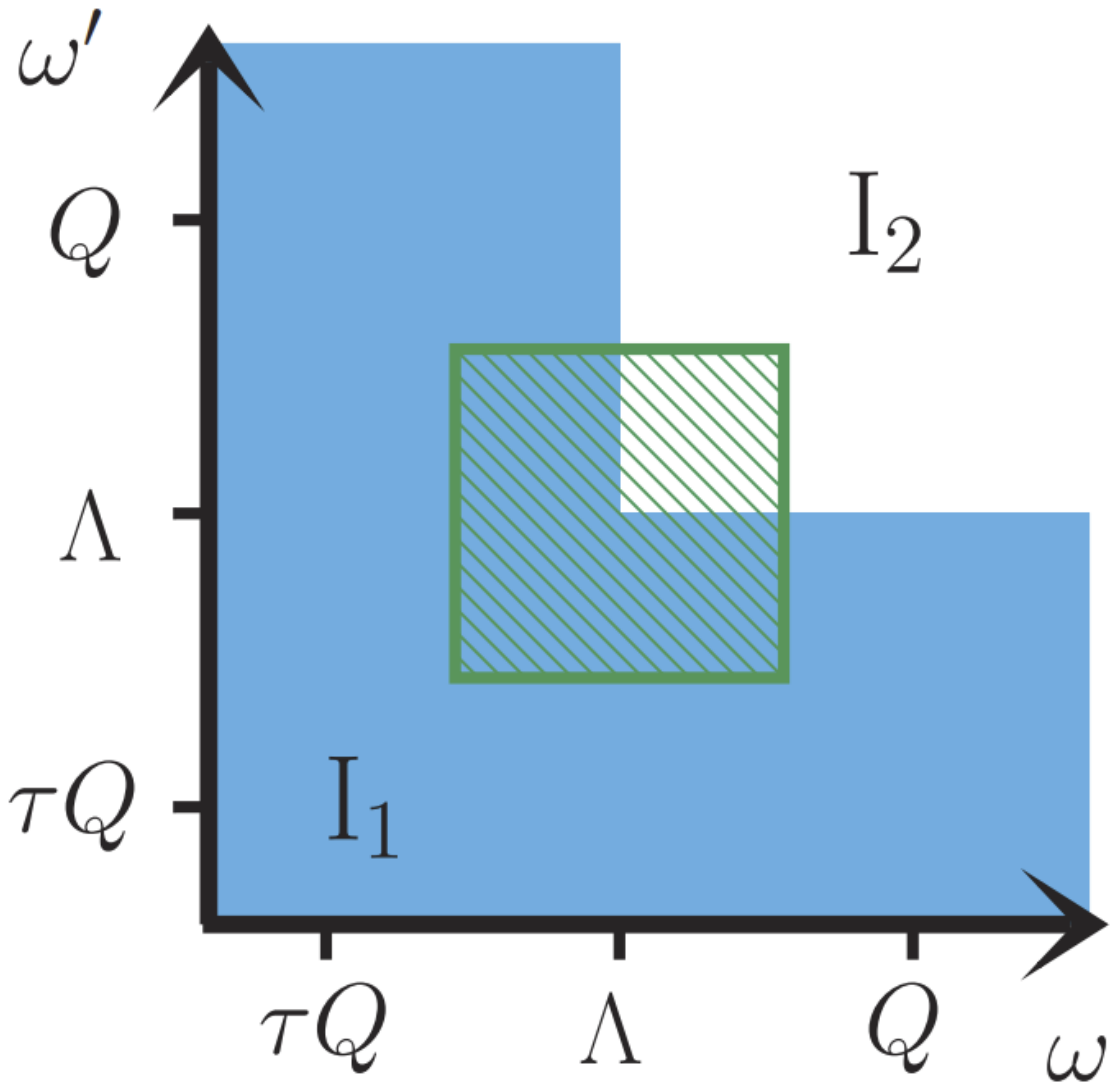}
\vspace{-1.0cm}
\caption{\label{fig:figure8}
	Visual representation of the regions $I_1$ and $I_2$, 
	used to construct subtraction terms as described in the
	text. In the overlap region in green the asymptotic 
	behaviour of the A- and B-type term must agree.
	Figure from \cite{Beneke:2022obx}.}
\end{center}
\end{figure}

Let us consider thrust in the two-jet limit: 
following \cite{Beneke:2022obx}, we consider 
the power-suppressed contribution given by 
the process (fig.~\ref{fig:figure5})
\be
e^{+}e^{-} \to \gamma^* \to 
[g]_c + [q\bar q]_{\bar c}.
\ee
Within SCET this process is given by two 
contributions \cite{Beneke:2022obx}: a 
``direct'' term (B-type) (first diagram 
on the left in fig.~\ref{fig:figure6}) 
and a time ordered product involving a 
soft quark emission (A-type) (two diagrams 
on the right in fig.~\ref{fig:figure6}): 
the former involves the matrix element
\begin{equation}
\langle X_c|\langle X_{\bar c}|\langle X_s|\,\bar{\chi}_{\bar c}(\bar{t}_1\nm)\Gamma_i^{\mu\nu}
\mathcal{A}_{c\perp\nu}(t\np)\chi_{\bar c}
(\bar{t}_2\nm) \, |0\rangle\,,
\label{eq:btypeME}
\end{equation}
with $i = 1,2$ representing two different
strings of Dirac matrices, while the latter 
stems from the matrix element
\begin{equation}
\langle X_c|\langle X_{\bar c}|\langle X_s| \int d^d x 
\,T\,[\bar{\chi}_c(t\np)\gamma_\perp^\mu\chi_{\bar c}
(\bar t\nm), i{\cal L}^{(1)\rm quark}_{c}(x)]\,|0\rangle\,.
\label{eq:atypetproduct}
\end{equation} 
Inserting these matrix elements in the 
corresponding cross section, one finds that 
the ``direct'' B-type term factorizes into 
a hard, (anti-)collinear and soft functions,
according to
\begin{eqnarray}
\frac{1}{\sigma_0}\frac{d\sigma}{dM_R^2 dM_L^2}
\bigg|_{\rm B-type} 
&\sim& 
	\frac{2 C_F}{Q^2} f(\eps) \int_0^\infty dl_+ dl_-\,
	\sum_{i,i'=1,2} \int dr dr'\,
	C^{\rm B1*}_{i'}(Q^2,r')C^{\rm B1}_{i}(Q^2,r) 
	\nonumber\\
	&&\hspace*{-2cm}\times\,
	\bigg\{\,
	\delta_{i i'}\,\mathcal{J}_{\bar{c}}^{q\bar{q}(8)}(M_R^2-Ql_+,r,r^\prime)
	+\,\ldots
	\bigg\}\,
	\mathcal{J}_{c}^{(g)}(M_L^2-Q l_-)\,S^{(g)}(l_+,l_-)\,,
	\label{eq:Btypefactformula}
\end{eqnarray}
where 
\begin{eqnarray}
f(\epsilon) = \bigg( \frac{Q^{2}}{4\pi} \bigg)^{-\epsilon} 
\frac{ (1-\epsilon)^2 \Gamma(1-\epsilon)}{ \Gamma(2- 2\epsilon)},
\end{eqnarray}
and the ellipses represent regular terms 
not important for our analysis below. Due 
to the fact that $C^{\rm B1}_{i}(r) \sim 1/r$, 
\eqn{eq:Btypefactformula} develops endpoint 
divergences when the quark ($r\to 0$) or the 
anti-quark ($r\to 1$) become soft: 
\be
\frac{1}{\sigma_0}\frac{d\sigma}{dM_R^2 dM_L^2}
\bigg|_{\rm B-type} 
\sim \int_0^1 dr \, \bigg[\frac{1}{r^{1+\eps}} 
+ \frac{1}{(1-r)^{1+\eps}}\bigg]\,.
\label{eq:BtypefactformulaEndpoint}
\ee
On the other hand, the A-type matrix element  
of \eqn{eq:atypetproduct} gives rise to the 
factorized cross section 
\begin{eqnarray}
\frac{1}{\sigma_0}\frac{d\sigma}{dM_R^2 dM_L^2}|_{\rm A-type} 
&\sim& \frac{2 C_F}{Q}\,
|C^{\rm A0}(Q^2)|^2 \int_0^\infty dl_+ dl_-\,
\int d\omega d\omega'\,
\mathcal{J}_{\bar c}^{(\bar q)}(M_R^2-Q l_+)
\nonumber\\
&&\hspace*{-0cm}\times\,
\bigg\{\,\mathcal{J}_{c}(M_{L}^{2}-Ql_-,\omega,\omega')
\,S_{\rm NLP}(l_+,l_-,\omega,\omega') + \ldots \bigg\}\,,
\label{eq:Atypefactformula}
\end{eqnarray}
which develops endpoint divergences when the 
soft quark or anti-quark become energetic 
($\omega \to \infty$):
\be
\frac{1}{\sigma_0}\frac{d\sigma}{dM_R^2 dM_L^2}|_{\rm A-type} 
\sim 2 \int_{M_R^2/Q}^{\infty} d\omega \, \frac{1}{\omega^{1+\eps}}\,.
\label{eq:AtypefactformulaEndpoint}
\ee
As for DIS, in the $r \to 0$ (or $r \to 1$)
limit the coefficients $C_i^{B1}$ become a 
two-scale object, and refactorize according 
to
\bea \label{eq:B1fact} \nn
C_1^{\rm B1}(Q^2,r) &\stackrel{r\to 0}{=}& C^{\rm A0}(Q^2)\times 
\frac{D^{\rm B1}(r Q^2)}{r}  +\mathcal{O}(r^0)\,, \\
C_2^{\rm B1}(Q^2,r) &\stackrel{r\to 1}{=}& -C^{\rm A0}(Q^2)\times 
\frac{D^{\rm B1}(\bar{r} Q^2)}{\bar r} +\mathcal{O}(\bar{r}^0)\,.
\eea
Furthermore, we expect that endpoint divergences 
should cancel when summing the A and B-type
contributions, which in turn implies that 
in the asymptotic limits $r,r' \to 0(1)$, 
$\omega, \omega' \to \infty$, the 
\emph{integrands} of the A- and B-type
terms should become identical. While it 
is easy to check that this is indeed what 
happens at lowest order in perturbation 
theory, in general this gives a series 
of refactorization conditions 
\cite{Beneke:2022obx}, which are 
summarized graphically in 
fig.~\ref{fig:figure7}: starting from 
the A-type term (upper left diagram 
in fig.~\ref{fig:figure7}), in the limit 
$\omega, \omega' \to \infty$ the soft
(anti)-quark becomes energetic, thus we 
get to the lower left diagram in 
fig.~\ref{fig:figure7}, provided that 
\cite{Beneke:2022obx}
\begin{eqnarray}
\mbox{(I)} &\hskip0.5cm 
\mathcal{J}_{c}\left(p^2,\omega,\omega'\right) = 
\mathcal{J}_{c}^{(g)}(p^2) \,
\displaystyle \frac{D^{\rm B1}(\omega Q)}{\omega}
\frac{{D^{\rm B1}}^*(\omega^\prime Q)}{\omega^\prime} 
+\mathcal{O}\!\left(\frac{1}{\omega^{(\prime)}}\right)\,,
\label{eq:EPfactI}
\end{eqnarray}
where the function $D^{\rm B1}(p^2)$ is 
the same as the one appearing in the 
factorization of the hard B1 operator 
coefficient \eqref{eq:B1fact}. On the 
other hand, starting from the B-type term 
(lower right diagram in fig.~\ref{fig:figure7}), 
in the limit $r,r' \to 0(1)$ the anticollinear
(anti-)quark becomes soft, thus we 
get to the upper right diagram in 
fig.~\ref{fig:figure7}, provided that 
\eqn{eq:B1fact} holds. At this point,
the requirement that the integrands 
of the A- and B-type terms should 
become identical in the asymptotic 
limits $r,r' \to 0(1)$, 
$\omega, \omega' \to \infty$, (i.e., 
that the lower left- and the upper 
right-diagram in fig.~\ref{fig:figure7}
should coincide),  gives the last 
refactorization condition: in 
Laplace space one has 
\cite{Beneke:2022obx}
\be
\mbox{(II)} \quad  
Q\,\widetilde{\mathcal{J}}^{(\bar{q})}_{\bar{c}}(s_R)\,
\widetilde{S}_{\rm NLP} \left(s_R,s_L,\omega,\omega'\right)
\Big|_{\omega^{(\prime)} \to\infty}  = 
\widetilde{\mathcal{J}}^{q\bar q(8)}_{\bar{c}}\left(s_R,r,r'\right)
\widetilde{S}^{(g)}(s_R,s_L)\Big|_{r^{(\prime)}=\omega^{(\prime)}/Q\to 0}\,,
\label{eq:EPfactII}
\ee
and the same identity holds 
with $r,r^\prime\to \bar{r},\bar{r}^\prime$. 
The constraint that in the asymptotic limits 
the A- and B-type terms must coincide 
provides also a method to deal with endpoint 
divergences. Let us define the asymptotic
limits of the various function by using a 
double-bracket notation: for instance, in 
functions of $\omega, \omega^\prime$, we rescale 
$\omega \to \kappa \omega$, $\omega^\prime 
\to \kappa \omega^\prime$ and take 
$\kappa\to \infty$. Then 
\begin{eqnarray}
&& \llbracket 
S_{\rm NLP} \left(l_+,l_-,\omega,\omega'\right)
\rrbracket \equiv S_{\rm NLP} 
\left(l_+,l_-,\omega,\omega'\right)|_{\mathcal{O}(\kappa^0)}, 
\label{eq:Sasym}\\
&&\llbracket \mathcal{J}_{c}(p^2,\omega,\omega^\prime)
\rrbracket \equiv 
\mathcal{J}_{c}(p^2,\omega,\omega^\prime)|_{\mathcal{O}(\kappa^{-2})}\,,
\end{eqnarray}
where $\delta(\omega-\omega')$ counts as $\kappa^{-1}$. 
A similar notation is used in functions of $r, r^\prime$
and we refer to \cite{Beneke:2022obx} for a precise 
definition. As discussed above, endpoint 
divergences arise in the asymptotic limits, where 
the A- and B-type terms take the factorized form 
given respectively in the left lower graph of 
fig.~\ref{fig:figure7}, and right upper graph 
of fig.~\ref{fig:figure7}: the cancellation of 
endpoint divergences requires these two limits 
to be identical, therefore we can introduce the 
integral
\begin{eqnarray}
&&\frac{2 C_F}{Q} \,f(\epsilon)\,|C^{\rm A0}(Q^2)|^2 
\widetilde{\mathcal{J}}^{(\bar q)}_{\bar{c}}(s_R) 
\widetilde{\mathcal{J}}_{c}^{(g)}(s_L) \nonumber\\
&&\times \int_0^\infty d\omega d\omega'\,
\frac{D^{\rm B1}(\omega Q)}{\omega}
\frac{D^{\rm B1*}(\omega'Q)}{\omega'} 
\left \llbracket \widetilde{S}_{\rm NLP}(s_R,s_L,\omega,\omega')\right 
\rrbracket,\qquad
\label{eq:scaleless_integral}
\end{eqnarray}
which is scaleless over the whole domain
and thus vanishes in $d$ dimensions, but it 
can be shown to reproduce respectively the 
endpoint divergences of the A- and B-type 
contribution, when splitting the integration 
into two domains $I_2$ and $I_1$, as represented 
in fig.~\ref{fig:figure8}\footnote{Let us notice 
that the splitting in fig.~\ref{fig:figure8}
is not unique. It is possible to 
split the integration domain differently, 
given that the endpoint divergences 
occur when both $\omega$ and $\omega'\to \infty$ 
or $r$ and $r' \to 0$ ($r$ and $r' \to 1$),
see \cite{Beneke:2022obx}.} 
\cite{Beneke:2022obx}. Thus it is 
possible to remove the endpoint 
divergences from both the A- and B-type 
contributions by subtracting the integrand
in \eqn{eq:scaleless_integral} integrated 
over region $I_2$ from the A-type term, 
and by subtracting \eqn{eq:scaleless_integral} 
integrated over region $I_1$ from the B-type 
term. The subtracted expressions are now 
separately endpoint-finite, but depend on 
$\Lambda$. However, as long as no approximations 
are made, the $\Lambda$ dependence cancels 
exactly between the two terms. After some 
elaboration \cite{Beneke:2022obx} the endpoint 
finite A-term can be written as
\begin{eqnarray}
\frac{1}{\sigma_0}\frac{\widetilde{d\sigma}}{ds_R ds_L}|_{\rm A-type} &=& 
\frac{2C_F}{Q} \,f(\epsilon)\,|C^{\rm A0}(Q^2)|^2 \,
\widetilde{\mathcal{J}}_{\bar c}^{(\bar q)}(s_R)
\,\int d\omega d\omega'\,\Big[1-\theta(\omega-\Lambda)\theta(\omega'-\Lambda)\Big]
\nonumber\\[0.0cm]
&&\hspace*{-2.5cm}\times\,
\bigg\{\,\widetilde{\mathcal{J}}_{c}(s_{L},\omega,\omega')
\,\widetilde{S}_{\rm NLP}(s_R,s_L,\omega,\omega')+\,\widetilde{\!\!\widehat{\mathcal{J}}}_{\!c}(s_{L},
\omega,\omega') 
\,\,\widetilde{\!\widehat{S}}_{\rm NLP}(s_R,s_L,\omega,\omega')\bigg\}\,,\qquad
\label{eq:Atype_subtracted}
\end{eqnarray}
where the equality signs hold up to corrections 
of $\mathcal{O}(1/(s_L\Lambda),1/(s_R\Lambda))$,
and the and B-type term takes the form
\begin{eqnarray}
\frac{1}{\sigma_0}\frac{\widetilde{d\sigma}}{ds_R ds_L}|_{
\scriptsize \begin{array}{l}
$\rm B--type$\\[-0.1cm] $i=i'=1$\end{array}} 
&=& 
\frac{2C_F}{Q^2}\,f(\epsilon) \,\widetilde{\mathcal{J}}_{c}^{(g)}(s_L)\,
\widetilde{S}^{(g)}(s_R,s_L)
\nonumber\\ 
&&\hspace*{-3cm}\times\,\bigg\{\int_0^1 dr dr' \,\big[1-\theta(\Lambda/Q-r)\theta(\Lambda/Q-r')\big]\,
C^{\rm B1*}_{1}(Q^2,r')C^{\rm B1}_{1}(Q^2,r) 
\,\widetilde{\mathcal{J}}_{\bar{c}}^{q\bar{q}(8)}(s_R,r,r^\prime)
\quad 
\nonumber \\ &&\hspace*{-2.5cm}-
\int_0^\infty dr dr' \,
\big[\theta(r-\Lambda/Q)\theta(\Lambda/Q-r')+
\theta(\Lambda/Q-r)\theta(r'-\Lambda/Q)\big]\,
\nonumber \\ &&\hspace*{-2cm}\times\,
\llbracket C^{\rm B1*}_{1} (Q^2,r')\rrbracket_{0}\, 
\llbracket C^{\rm B1}_{1}(Q^2,r) \rrbracket_{0} \, \llbracket \widetilde{\mathcal{J}}_{\bar{c}}^{q\bar{q}(8)}(s_R,r,r^\prime)\rrbracket_0\,
\bigg\}\,,
\label{eq:Btype_subtracted3}
\end{eqnarray}
up to corrections of $\mathcal{O}(\Lambda/Q)$. 
It is now possible to use these expressions to 
develop the resummation of large logarithms 
with standard methods, and we refer to 
\cite{Beneke:2022obx} for a detailed 
discussion of such derivation. 

\section{NLP NNLO in Drell-Yan near threshold}

The analysis of off-diagonal thrust 
has allowed us to fully appreciate the 
nature of endpoint divergences: these 
are indeed an artifact of the effective
field theory, which arise due to how
the original (phase space or loop) 
integrations in QCD are split 
among the different regions of the 
effective theory. When all momentum 
regions are correctly taken into account, 
(A- and B-type terms in case of off-diagonal 
thrust), endpoint divergences cancel. This 
allows one to devise the construction of 
subtraction terms such as to make the 
individual contributions finite. Formally, 
the finite factorization formulas such as 
\eqns{eq:Atype_subtracted}{eq:Btype_subtracted3}
are expected to be valid at all orders in 
perturbation theory and at any logarithmic 
accuracy. However, for off-diagonal thrust 
the explicit construction has been obtained 
at LL accuracy. In order to go beyond this 
logarithmic order, in general one needs 
to calculate the factorized matrix elements
beyond leading order in perturbation theory.
In turn, this would allow one to check 
explicitly that the refactorization conditions 
such as those in \eqns{eq:EPfactI}{eq:EPfactII} 
hold to higher order in perturbation theory. 

With this goal in mind, a set of papers
\cite{Beneke:2019oqx,Broggio:2021fnr,Broggio:2023pbu}
have been dedicated to calculate the 
full set of collinear and soft functions 
necessary to reproduce the Drell-Yan
invariant mass distribution at NLP 
near threshold, up to NNLO in 
perturbation theory. In particular, 
this requires the calculation of the 
collinear functions up to one loop 
and the soft function up to two loops.
Let us focus for simplicity on the off
diagonal $\bar q g$ channel. In 
\cite{Broggio:2023pbu} the 
factorization theorem has been 
obtained formally at all subleading
power. Writing the invariant mass 
distribution as
\begin{equation}
\frac{d\sigma_{\rm DY}}{dQ^2} = \sigma_0 
\sum_{a,b} \int_{\tau}^1\, \frac{dz}{z}\, 
{\cal L}_{ab}\bigg(\frac{\tau}{z}\bigg)\, \Delta_{ab}(z) 
+ \mathcal{O}\left(\frac{\Lambda}{Q}\right), \qquad 
\sigma_0 = \frac{4\pi \alpha_{\rm em}^2}{3 N_c Q^2 s},
\label{eq:dsigsqDelta}
\end{equation}
where the parton luminosity function 
${\cal L}_{ab}(y)$ is defined as 
\be
{\cal L}_{ab}(y) = \int_y^1\frac{dx}{x} \, f_{a/A}(x) \, 
f_{b/B}\bigg(\frac{y}{x}\bigg),
\ee
up to NLP the partonic cross section 
factorizes as follows: 
\be\label{eq:QGnlpfact}
\Delta_{g\bar{q}}|_{\rm NLP}(z) = 
8 H(Q^2)\,\int {d\omega}\, d\omega' \,  
G^*_{\xi q}(x_an_+p_A;\omega') \,
G_{\xi q}(x_an_+p_A;\omega) 
\, S(\Omega,\omega, \omega') \,,
\ee
where $G_{\xi q}$, $G^*_{\xi q}$
are collinear function appearing 
respectively in the amplitude
and complex conjugate amplitude,
$H(Q^2) = |C^{A0}(Q^2)|^2$ is the 
hard function, and $S(\Omega,\omega, \omega')$
the soft function. The collinear
function coincides with 
the function $D^{B1}$ in 
\eqns{eq:B1fact}{eq:EPfactI}. 
Indeed, this collinear matrix 
element appears to be a ``universal''
function appearing in the context 
of several factorization theorems 
at NLP; for instance, it appears 
also in the context of 
$H \to \gamma\gamma$, \cite{Liu:2019oav}, 
and it has been calculated up to 
two loops in \cite{Liu:2021mac}. 
At one loop it reads 
\bea\label{eq:qgoneloopcollfuncscalar} 
{G}^{(1)}_{\xi q}\left(n_+p; \omega \right) 
&=& -\, \frac{\alpha_s}{4\pi} \big(C_F - C_A \big)
\left(\frac{n_+p\, \omega}{\mu^2}\right)^{-\epsilon} 
\frac{2-4\eps-\eps^2}{2\epsilon^2}
\frac{e^{\epsilon \gamma_E}\Gamma[1+\epsilon]
\Gamma[{1-\epsilon}]^2}{\Gamma[{2-2\epsilon}]}  \\ \nn
&=&-\,\frac{\alpha_s}{4\pi} \big(C_F - C_A \big)
\bigg[\frac{1}{\epsilon ^2} -\frac{1}{\epsilon}
\ln \left(\frac{n_+p \,\omega}{\mu^2}\right) -\,\frac{1}{2} -\frac{\pi^2}{12}
+ \frac{1}{2}\ln^2\left(\frac{n_+p\, \omega_1}{\mu^2}\right) 
+\mathcal{O}(\epsilon) \bigg]\,.
\end{eqnarray}
The soft function at one loop 
reads 
\be\label{eq:nlo-matrix} 
S_{g\bar{q}}^{(1)}(\Omega,\omega,\omega')
= \frac{\alpha_s \, T_F}{4\pi}  
\frac{e^{\epsilon \gamma_E}}{
\Gamma[1-\epsilon]} \frac{1}{\omega} 
\bigg(\frac{\mu^{2}}{\omega\,
(\Omega-\omega)}\bigg)^{\eps}
\delta(\omega- \omega') \,
\theta(\Omega-\omega) \theta(\omega),
\ee
where the color factor $T_F$ 
is given in terms of the relation 
${\rm Tr}[\T^A \T^B] = T_F\,\delta^{AB} 
= \delta^{AB}/2$. The two loop contribution
is quite involved. It includes a 
virtual-real and a real-real contributions
\be
S^{(2)}_{g\bar{q}}(\Omega,\omega,\omega') 
= S^{(2)1r1v}_{g\bar{q}}(\Omega,\omega,\omega') 
+ S^{(2)2r0v}_{g\bar{q}}(\Omega,\omega,\omega'),
\ee
which individually read
\bea \nonumber
\label{soft-function-momentum-1r1v-NNLOd}
S^{(2)1r1v}_{g\bar{q}}(\Omega,\omega,\omega')  
&=& \frac{\alpha_s^2\, T_F}{(4\pi)^2} (2C_F-C_A) 
\frac{e^{2\eps \gamma_E}\,\Gamma[1+\eps]}{\eps\,
	\Gamma[1-\eps]} \\ \nn
&&\hspace{-3.0cm} \times \,
{\rm Re}\bigg\{ \frac{1}{(-\omega)\omega'}
\bigg[\frac{\omega+\omega'}{\omega'}
\, _2F_1 \bigg(1,1+\eps,1-\eps,\frac{\omega}{\omega'}\bigg)
- 1 \bigg] \bigg(\frac{\mu^4}{(-\omega) 
	\omega'(\Omega - \omega')^2}\bigg)^{\eps}
\theta(-\omega)   \\
&&\hspace{-2.7cm}+\, \frac{2(\omega+\omega')}{\omega\omega'(\omega'-\omega)} 
\bigg(\frac{\mu^4}{(\omega'-\omega)^2(\Omega - \omega')^2}\bigg)^{\eps}
\frac{\Gamma[1-\eps]^2}{\Gamma[1-2\eps]}\theta(\omega'-\omega)
\bigg\} \theta(\omega')\theta(\Omega-\omega'),
\eea
and
\begin{eqnarray} 
	\label{soft-function-momentum-2realc} \nonumber 
	S^{(2)2r0v}_{g\bar{q}}(\Omega,\omega,\omega') 
	&=& \frac{\alpha_s^2 T_F}{(4\pi)^2} 
	\bigg\{ C_F \frac{e^{2 \epsilon \gamma_E} 
		\Gamma[1-\eps]}{\eps^2} \frac{1}{\omega}
	\bigg[\frac{4}{\Gamma[1-3\eps]}
	\bigg(\frac{\mu^4}{\omega(\Omega-\omega)^3}\bigg)^{\eps} 
	\\ \nonumber
	&& \hspace{0.0cm}
	+\,\frac{(4-\eps)\Gamma[2-\eps]}{(1-2\eps)\Gamma[1-2\eps]^2}
	\bigg(\frac{\mu^4}{\omega^2(\Omega-\omega)^2}\bigg)^{\eps}
	\bigg] \delta(\omega - \omega')
	\theta(\Omega-\omega)\theta(\omega) \\ 
	&& \hspace{-1.0cm}
	+\,\left(C_A - 2 C_F\right) 
	\frac{2e^{2 \epsilon \gamma_E}}{
		\eps \Gamma[1-2\eps]} \frac{\omega
		+\omega'}{\omega \omega'(\omega'-\omega)}
	\bigg(\frac{\mu^4}{\omega(\omega'-\omega)
		(\Omega-\omega')^2}\bigg)^{\eps} \\ \nonumber 
	&& \hspace{0.0cm} \times \, 
	\bigg[ \,_2F_1\Big(1,-\epsilon,1-\epsilon, 
	\frac{\omega}{\omega-\omega'} \Big) -1 \bigg] 
	\theta(\omega)\theta(\omega')
	\theta(\omega'-\omega) \theta(\Omega-\omega')\bigg\}.
\end{eqnarray}
With these results at hand one can now
study the structure of endpoint divergences 
in the asymptotic limits $\omega,\omega' \to 0$, 
and we refer to \cite{Broggio:2023pbu} for 
further details. 

\section{Outlook}

The resummation of large logarithms at NLP
poses interesting theoretical challenges. 
Within an effective field theory approach 
based on SCET it is possible to systematically 
factorize the effect of soft and collinear 
radiation in physical observables. In this 
talk we have discussed the derivation of 
factorization theorems for scattering 
processes such as Drell-Yan and deep 
inelastic scattering near threshold, 
and thrust in the two-jet limit. In 
general, once bare factorization 
theorems have been derived, the 
subsequent step of obtaining the 
resummation of large logarithms 
by means of a RGE approach is made 
nontrivial by the appearance of endpoint 
divergences \cite{Beneke:2019oqx,Beneke:2020ibj}. 
It has been shown that these are an artifact 
of the effective theory. Endpoint divergences
cancel among terms in the factorization theorems, 
once all contributions are correctly taken into 
account. It is then possible to devise a 
subtraction procedure, which makes the 
individual contributions finite, thus 
allowing the application of standard 
RGE procedure. This approach has been 
fully developed to obtain the resummation 
of large logarithms in off-diagonal 
``gluon thrust'' at LL accuracy 
\cite{Beneke:2022obx}. In 
order to extend resummation at NLP to 
higher logarithmic accuracy more data 
is needed. To this end one needs to 
calculate collinear and soft functions 
appearing in the factorization theorems 
at higher order in perturbation theory.
This program has been completed for 
Drell-Yan near threshold in a series of
papers, \cite{Beneke:2019oqx,Broggio:2021fnr,Broggio:2023pbu}, 
where all collinear and soft functions
have been calculated respectively at one 
and two loops in perturbation theory.

\section*{Acknowledgment}

This work has been partly supported by 
Fellini - Fellowship for Innovation 
at INFN, funded by the European Union's 
Horizon 2020 research programme under 
the Marie Sk\l{}odowska-Curie Cofund 
Action, grant agreement no. 754496.

\bibliography{skeleton}

\end{document}